\documentclass[preprintnumbers,nofootinbib,noshowpacs,eqsecnum,prd,superscriptaddress,letterpaper]{revtex4}

\usepackage{graphicx}
\usepackage{amsmath,amssymb}
\usepackage{url}
\usepackage{ulem}

\makeatletter 
\renewcommand{\thesection}{\arabic{section}}
\renewcommand{\p@subsection}{}

\makeatother

\newcommand{\gesim}{\,{_{\textstyle >}\atop^{\textstyle\sim}}\,}

\newcommand{\gl}[1]{\eqref{#1}}

\begin{document}

\title{Constraining the Intrinsic Structure of Top-Quarks}

\author{C.~Englert}
\affiliation{Institute for Particle Physics Phenomenology, Department
    of Physics, Durham University, DH1 3LE, United Kingdom}
\author{A.~Freitas}
\affiliation{Pittsburgh Particle-physics Astro-physics \& Cosmology
    Center (PITT-PACC),  Department of Physics \& Astronomy,
    University of Pittsburgh, Pittsburgh, PA 15260, USA}
\author{M.~Spira}
\affiliation{Paul Scherrer Institut, W\"urenlingen und Villigen,
    CH-5232 Villigen PSI, Switzerland}
\author{P.M.~Zerwas}
\affiliation{Deutsches Elektronen-Synchrotron DESY, D-22603 Hamburg,
    Germany}

\begin{abstract} {\sl \noindent The basic structure of top-quarks as
    spin-1/2 particles is characterized by the radius $R_t$ and the
    intrinsic magnetic dipole moment $\kappa_t$, both individually
    associated with gauge interactions. They are predicted to be zero
    in pointlike theories as the Standard Model. We derive upper
    limits of these parameters in the color sector from cross sections
    measured at Tevatron and LHC in top-pair production $p{\bar{p}}/pp
    \to t{\bar{t}}$, and we predict improved limits expected from LHC
    in the future, especially for analyses exploiting boosted top
    final states. An additional method for measuring the intrinsic
    parameters is based on $t \bar{t} + jet$ final states.  }
\end{abstract}

\preprint{IPPP/12/66}
\preprint{DCPT/12/132}
\preprint{PSI--PR--12--05}

\maketitle

\setcounter{section}{1}
\setcounter{equation}{0}
\noindent {\bf 1. Basic Set-up.} The top-quark is the heaviest particle in the
Standard Model (SM), even if the Higgs particle is included as a
contender. This observation led to many approaches in which the
top-quark plays the role of portal to physics beyond the Standard
Model, see {\it e.g.} Refs.\cite{BSM,Hol}. Scales characterizing the
novel interactions in which the top-quark is identified with the
crucial source field, may be realized not far beyond the TeV size. As
a consequence, the top-quark may be endowed with intrinsic structure
at the TeV scale.  This should be contrasted with the pointlike
character of all fundamental fields within the Standard Model,
extending up to scales close to the Planck scale for low Higgs mass.\\

The basic non-pointlike structure will manifest itself in a non-zero
{\it radius} $R_t$ and a non-zero {\it anomalous magnetic dipole
  moment} $\kappa_t$ in $CP$-invariant scenarios, probed in
interactions with gauge fields \cite{Drell}. Due to the high energy
available, the LHC will enable us to probe the intrinsic top-quark
structure in the colored sector at an unprecedented
level~\cite{magmom,hioki,anomcpl,anomcpl2,CT}. Non-pointlike
interactions with the gluon field{\footnote{The massless gluon gauge
    field is assumed intrinsically pointlike in the present
    analysis. This assumption can be removed, see
    Ref.~\cite{Fritzsch}, at the expense of increasing
    complexity. Non-pointlike structures of the weak current remain
    non-effective as long as the top decay is treated inclusively with
    $BR(t \to bW)$ very close to unity.}}  modify the color quark
current to \cite{Drell}
\begin{equation} 
  \label{eq:curr}
  {\mathcal{J}}_\mu = F_t\, \gamma_\mu + i\,
  \frac{\kappa_t}{2 m_t} \, \sigma_{\mu\nu} Q^\nu \,.
\end{equation} 
The current incorporates the form factor
\begin{equation}
F_t = 1 + \dfrac{1}{6} R_t^2 Q^2 \,,
\end{equation}
with the top-quark radius $R_t$ related by
\begin{equation}
  R_t = \sqrt{6}/\Lambda_\ast
  \label{eq:Rlam}
\end{equation}
to the new scale parameter $\Lambda_\ast$, and the anomalous
chromo-magnetic dipole moment $\kappa_t$ [beyond the loop value
\cite{Schwinger}]. To protect fermion masses from acquiring large
values, the theory is generally assumed chiral \cite{Brod}, and the
breaking of the chiral symmetry by anomalous magnetic moments is
suppressed by two powers of the scale $\Lambda_\ast$, in the simplest
possible realization:
\begin{equation} 
  \kappa_t = \rho m_t^2/\Lambda_\ast^2    \,,
\label{eq:klam}
\end{equation}
where $|\rho|$ is an ${\cal O}(1)$ number.  The quadratic
$\Lambda_\ast$ dependence of $\kappa_t$ is effectively equivalent to
the scaling of the form factor. The quadratic dependence in the heavy
quark mass singles out the top-quark as unique particle for which
$\kappa_t$ may be accessible experimentally, in contrast to much less
sensitive light quarks or leptons. Assuming $\Lambda_\ast$ to be of
order 1 TeV and beyond, compatible with bounds on contact interactions
from Tevatron and LHC \cite{contact},
$\kappa_t$ could be expected at the level of several per-cent.  \\

Both the anomalous parameters, color radius and color magnetic dipole
moment, can be introduced through effective Lagrangians \cite{BWCT} in
an SU(3)$_c$ gauge-invariant and parity-even
form{\footnote{Electroweak gauge invariance can be ensured by
    expanding the Lagrangians to the complete third generation and
    incorporating the Higgs field \cite{BWCT}.}}:
\begin{eqnarray} 
  \label{eq:interactions} 
  \mathcal{L}_R &=& -g_s
  \frac{R_t^2}{6} \; \bar{t} \gamma^\mu \, \mathcal{G}_{\mu\nu} \, D^\nu
  \, t + {\rm h.c.},  \\ \label{eq:interactions2} \mathcal{L}_\kappa &=&
  g_s \,\frac{\kappa_t}{4 m_t}\; \bar{t} \sigma^{\mu\nu} \,
  \mathcal{G}_{\mu\nu} \, t                                  \,,
\end{eqnarray}
with the gluon field $\mathcal{G}_\mu$, in octet matrix notation, and
the gluon field strength $\mathcal{G}_{\mu\nu} = D_\nu \mathcal{G}_\mu
- D_\mu \mathcal{G}_\nu $, while $D^\nu = \partial^\nu + i g_s
\,\mathcal{G}^\nu$ denotes the covariant derivative of QCD. Besides
the components generating the anomalous top color current, the
Lagrangians are complemented by additional two-gluon and three-gluon
top interactions, as demanded by gauge invariance. The effective
Lagrangians unambiguously translate the anomalous parameters
from scattering to annihilation processes. \\

The classical method for studying radius and anomalous magnetic dipole
moment of the top quark is given by the elastic Rutherford-type
scattering of a top quark $t$ with a light quark $q$ [taken pointlike
in the present scenario], which is mediated by the exchange of a gluon
in $qt \to qt$. Rutherford-type scattering is also embedded in the
process $gq \to t \bar{t} q$. At very high energies, gluon partons in
the protons split into beams of long-lived top-quark pairs traveling
parallel to the gluon momentum. Thus, the events of the $t \bar{t} q$
process, characterized by a forward moving $t$-quark plus a $\bar{t}
q$-pair, with the two partons in the pair balanced in transverse
momentum, signal Rutherford $qt$ scattering. [Elastic gluon-top
scattering is independent of the radius $R_t$ and cannot be
exploited.] \label{ruth}\\[2mm]

\setcounter{section}{2}
\setcounter{equation}{0}
\noindent {\bf 2. Theoretical groundwork.} We will analyze the total 
cross sections for the production of top-quark pairs
\begin{equation}
p \bar{p} /pp \to q\bar{q},gg \to t \bar{t} 
\end{equation}
at Tevatron and LHC for deriving limits on the color radius $R_t$, the
anomalous chromo-magnetic dipole moment $\kappa_t$ and the
$\Lambda_\ast$ parameter in practice. Additional constraints can be
derived from the angular dependence of the top-quarks, and the
correlations between longitudinal spin components of $t$ and $\bar{t}$
\cite{Bern}, which can be measured unperturbed by fragmentation due to
the short top lifetime \cite{Dokshitzer}. Related analyses have been
discussed in Refs.~\cite{Degrande:2010kt,weiler,hesari}. \\

We will assume that the non-pointlike contributions to the observables
are small and, correspondingly, we will expand the observables
linearly in the analytic formulae. In fact, anomalous chromo-magnetic
dipole moment and chromo-radius are the first terms of a multipole
expansion including scale parameters beyond the Standard Model. The
systematic expansion would continue with higher-order moments the
quadratic terms in $R_t^2$ and $\kappa_t$ would compete with. An
analysis of these contributions is beyond
the scope of the present letter. \\

The hadron cross sections are built up by the incoherent superposition
of quark-antiquark annihilation and gluon fusion to top-antitop
pairs. Quark-antiquark annihilation is mediated only by $s$-channel
gluon exchange\footnote{We neglect electroweak interactions in the
  following.}, gluon fusion by
$s$-channel gluon and $t,u$-channel top exchanges. \\

The anomalous terms of the independent cross sections at the parton
level can be summarized as follows [see also references quoted above],
using $\beta = \sqrt{1-4m_t^2/s}$, where $s$ is the partonic
center-of-mass energy:\\[-2mm]

\noindent
{\uline{\it quark-antiquark annihilation}}:  
\begin{align}
  \label{eq:1}
  &\dfrac{\Delta\sigma}{\sigma_B} = \frac{s}{3} R_t^2 + 
   \dfrac{6 \kappa_t}{3 - \beta^2}  
\end{align}
\begin{align}
  &\dfrac{\Delta d\sigma/d\cos\theta}{d\sigma_B/d\cos\theta} = 
   \frac{s}{3} R_t^2 + \dfrac{4 \kappa_t}{2 - \beta^2(1-\cos^2\theta)} \\[2mm]
  &\dfrac{\{t_R \bar{t}_R+t_L \bar{t}_L\}-\{t_R \bar{t}_L+t_L \bar{t}_R\}}
  {\{t_R \bar{t}_R+t_L \bar{t}_L\}+\{t_R \bar{t}_L+t_L \bar{t}_R\}} = 
   -\frac{1+\beta^2}{3-\beta^2} + \frac{8\beta^2}{(3-\beta^2)^2}\kappa_t \,. 
\end{align}

\noindent
{\uline{\it gluon fusion}}:                                                                                                                                              
\begin{align}  
  \label{eq:2}
  &\dfrac{\Delta\sigma}{\sigma_B} = \frac{(36\beta
    -64\tanh^{-1}\beta)\kappa_t}{
    \beta(59-31 \beta^2) - 2(33-18\beta^2+\beta^4)\tanh^{-1}\beta}  \\[2mm]
  &\dfrac{\Delta d\sigma/d\cos\theta}{d\sigma_B/d\cos\theta} =
  \frac{4(1-\beta^2\cos^2\theta)\kappa_t}{2-\beta^4-
    (1-\beta^2+\beta^2\cos^2\theta)^2}\\[2mm]
  &\dfrac{\{t_R \bar{t}_R+t_L \bar{t}_L\}-\{t_R \bar{t}_L+t_L
    \bar{t}_R\}} {\{t_R \bar{t}_R+t_L \bar{t}_L\}+\{t_R \bar{t}_L+t_L
    \bar{t}_R\}} = -\frac{\beta(66-37 \beta^2+31\beta^4) -
    2(33-33\beta^2+17\beta^4-\beta^6)
    \tanh^{-1}\beta}{\beta^2[\beta(59-31 \beta^2) -
    2(33-18\beta^2+\beta^4)\tanh^{-1}\beta]} \\
  &\hspace{8em} - \frac{4[\beta(33+11\beta^2) -
    (33-\beta^4)\tanh^{-1}\beta] [\beta(41-31\beta^2) -
    2(17-18\beta^2+\beta^4)\tanh^{-1}\beta]} {\beta^2[\beta(59-31
    \beta^2) - 2(33-18\beta^2+\beta^4)\tanh^{-1}\beta]^2}\kappa_t
  \nonumber \,,
\end{align}
in agreement, wherever overlapping, with {\it e.g.}
\cite{haberl,Degrande:2010kt}. Other helicity asymmetries are related
by $\mathcal{P}$ and $\mathcal{C}$ invariance. The effective top
current Eq.~\gl{eq:curr}, generates the same dependence on the
anomalous parameters in the $q\bar q$ amplitude, so that the top
radius and the anomalous magnetic moment $R_t,\kappa_t$ can indeed
be interpreted as gauge-invariant characteristics of the top quark.\\

The $q \bar{q}$ annihilation channel is modified by both the radius
and the chromo-magnetic moment. By contrast, gluon fusion does not
depend on the radius to leading order -- reminiscent of the Thomson
cross section in QED -- but only on the anomalous magnetic moment (see
Appendix and {\it e.g.} Ref.~\cite{Degrande:2010kt}). Since top
production at the Tevatron is driven by $q \bar{q}$ collisions, both
parameters can in principle be determined in top measurements at this
collider.  On the other hand, the LHC, where gluon fusion is by far
the dominant inclusive top channel, is highly sensitive to the value
of the color anomalous magnetic dipole moment, leading to large bounds
on the scale parameter $\Lambda_\ast$. However, at the expense of
reduced cross sections, the relative weight of the $q \bar{q}$ channel
can be increased by the production of boosted top events at LHC, and
$t \bar{t}$ production in this configuration becomes also sensitive
to the radius. \\

Naturally assuming universality for the light quarks $q = u,d$,
the bounds on $R_t$ can be transcribed easily to the scale of the
standard color-octet vectorial contact interactions
${\mathcal{L}}_{\rm ct} = g_{\rm ct}^2/{\Lambda}_{\rm ct}^2 \,
(\bar{q} \gamma^\mu T^A\, q)(\bar{t} \gamma_\mu T^A\, t)$, where
$T^A=\lambda^A/2$ denote the SU(3) generators of QCD, expressed by the
Gell-Mann matrices $\lambda^A$. After inserting the effective contact
coupling $g_{\rm ct}^2 = 4\pi$ of the two quark currents, as generally
defined, the contact scale $\Lambda_{\rm ct}$ is related to the
compositeness scale $\Lambda_{\ast}$ by
\begin{equation}
  {\Lambda}_{\rm ct} \sim  1/ \sqrt{\alpha_s} \; {\Lambda}_{\ast}  \,.
\label{Eq:ct}
\end{equation}  
The different coupling strengths boost the contact scale to a value
half an order of magnitude above the compositeness scale.  Current
constraints limit the octet contact scale to $\Lambda_{\rm ct} \gesim
2.8$~TeV, see below and Ref.\cite{Degrande:2010kt}.  The singlet
contact scale of general chiral quark interactions has been
constrained to $\geq$ 3.4 TeV at LHC \cite{contact}; this bound may be
compared with $\sqrt{3/\sqrt{2}}\, \Lambda_{ct} \simeq 4.1$ TeV for
top interactions if the singlet energy density is identified,
hypothetically, with the octet density.\\[2mm]

\setcounter{section}{3}
\setcounter{equation}{0}
\noindent {\bf 3. Numerical Evaluation.} The determination of the anomalous 
parameters by three independent measurements of cross sections at three different
energies and different superpositions of the parton subprocesses at
Tevatron and LHC is over-constrained. Leaving the exhaustive
evaluation to experimental analyses proper we focus in this
theoretical study on the total cross sections at Tevatron and LHC.
Combining the cross sections of both colliders the different weight of
$q \bar{q}$ and $gg$ events allows us to separate the parameters $R_t$
and $\kappa_t$. We will also investigate the cross section for boosted
final-state tops, which are well-accessible at the LHC with 14 TeV
center-of-mass energy, again initiated by $q \bar{q}$ and $gg$ parton
compositions different from inclusive cross sections.  These
experimental observables are well documented by both
the collaborations at the two colliders \cite{Tevatron,LHC,atlasboost}.\\

The $t\bar t +X$ cross section follows from the modified Born-level
$t\bar t$ amplitudes
${\cal{M}}={\cal{M}}_{\text{SM}}+{\cal{M}}(\kappa_t,R_t)$ for the
partonic subprocesses $ab=q\bar q,gg$, where $q$ denotes the light
quark flavors, so that

\begin{equation}
  \begin{split}
    \label{eq:mcimp}
    \Delta\sigma &= \sum_{ab \in \{q\bar q,gg\}} \,\iiint {\text{d}} x_1\,
    {\text{d}} x_2\,  
    {\text{d}} {\text{LIPS}} \, f_a(x_2,\mu_F^2)  f_b(x_2,\mu_F^2)\,
    \left\{ |{\cal{M}}_{ab} |^2 -
      |{\cal{M}}_{\text{SM}\,ab}|^2 \right\} \\
    &= \sum_{ab \in \{q\bar q,gg\}} \,\iiint {\text{d}} x_1\,
    {\text{d}} x_2\,  
    {\text{d}} {\text{LIPS}} \, f_a(x_2,\mu_F^2)  f_b(x_2,\mu_F^2)\,
    2{\text{Re}}\left\{
      {\cal{M}}_{\text{SM}\,ab}^\ast{\cal{M}}_{ab}(\kappa_t,R_t)
    \right\}+{\cal{O}}\left({1\over \Lambda_\ast^4}\right)\,.
  \end{split}
\end{equation}
For the remainder of this analysis we choose the CTEQ6L1 parton
distribution set \cite{cteq}. \\

We have implemented the parton-level cross section of
Eq.~\gl{eq:mcimp} in a fully flexible numerical program based on the
{\sc{Vbfnlo}} framework \cite{vbfnlo}. The calculation of the matrix
elements is performed with a set of custom-built {\sc{Helas}} routines
\cite{helas} which facilitate the numerical evaluation of the three-
and four-point contributions of Eq.~\gl{eq:interactions}. The
$R_t$-dependent terms in the on-shell gluon-induced subprocess drop
out, as discussed earlier. This cancellation persists in the
gluon-initiated subprocess of
$\stackrel{}{p}\stackrel{\text{\tiny(}-\text{\tiny)}}{p} \to t\bar
t+{\rm{jet}}$ production, which incorporates off-shell gluons by
emitting an additional jet (see below). However, this process is still
worth studying since the quark-gluon initiated channel first enters at
this order and thus offers new ways of probing $R_t$.  The numerical
implementation for
$\stackrel{}{p}\stackrel{\text{\tiny(}-\text{\tiny)}}{p} \to t\bar
t+{\rm{jet}}$ is set up analogously to
$\stackrel{}{p}\stackrel{\text{\tiny(}-\text{\tiny)}}{p} \to t\bar t$,
supplementing the relevant five-point interactions following from
Eq.~\gl{eq:mcimp}. We have checked all contributing matrix elements
for gauge invariance and we have validated our phase space integration
against {\sc{MadEvent}} \cite{madevent} and
{\sc{Sherpa}} \cite{sherpa}. \\[2mm]

The upper limits $\Delta\sigma(t\bar t +X)$ as generated{\footnote{The
    corresponding code for $\Delta\sigma$ is available upon request
    from the authors.}} by the anomalous top parameters, color radius
$R_t$ and magnetic dipole moment $\kappa_t$, are identified with the
difference between the measured and the theoretically predicted SM
cross sections, both including errors (for details see below):
\begin{align}
  \label{eq:deltasig}
  \hbox{Tevatron (CDF)~\cite{Tevatron}:}\quad &\sigma(t\bar t+X) = 7.5
  \pm
  0.31 \,{\rm{(stat)}} \pm 0.34 \, {\rm{(syst)}} \pm 0.15 \,  {\rm{(lumi)}} ~{\rm{pb}} \\
  \hbox{LHC, $\sqrt{s}=7$ TeV~\cite{LHC}:}\quad & \sigma(t\bar t+X) =
  177\pm 3 \,{\rm{(stat)}}\;^{+8}_{-7}\,{\rm{(syst)}} \pm 7
  \,{\rm{(lumi)}}~{\rm{pb}}\,.
\end{align}
We have adopted the theoretical expectations of Ref.~\cite{moch}
\begin{align}
  \label{eq:deltasigtev}
  \hbox{Tevatron:}\quad &\sigma(t\bar t+X) = 7.13\;
  ^{+0.30}_{-0.40} \,{\rm{(scale)}} \;
  ^{+0.17}_{-0.12}\,{\rm{(pdf)}}~{\rm{pb}} \\
  \label{eq:deltasiglhc7}
  \hbox{LHC, $\sqrt{s}=7$ TeV:}\quad & \sigma(t\bar t+X)  =
  164.3 \; ^{+3.3}_{-9.2}\,{\rm{(scale)}}
  \;^{+4.4}_{-4.5}\,{\rm{(pdf)}}~{\rm{pb}}\\
  \label{eq:deltasiglhc14}
  \hbox{LHC, $\sqrt{s}=14$ TeV:}\quad & \sigma(t\bar
  t+X) = 908.3 \;
  ^{+9.8}_{-40.5}\,{\rm{(scale)}}\;^{+15.2}_{-16.7}\,{\rm{(pdf)}}~
  {\rm{pb}}
\end{align}
as representative figures of the inclusive $t \bar t$ cross sections
\cite{ttTH,ttTH2,ttTH3}. It has been shown in Refs.~\cite{ttTH2,ttTH3}
that the perturbative evolution up to the full NNLO precision result
for $t\bar t$ production at the Tevatron reduces the renormalization
scale uncertainty by ${\cal{O}}{(30\%)}$ and a similar improvement is
expected for LHC predictions. We include the theoretical uncertainty
due to variations of the renormalization scale and errors of the
parton densities by adding it to the previously mentioned experimental
error in quadrature; the differences of the theoretical
and experimentally expected mean values are added equivalently. \\

\begin{figure}[!t]
  \includegraphics[height=0.32\textwidth]{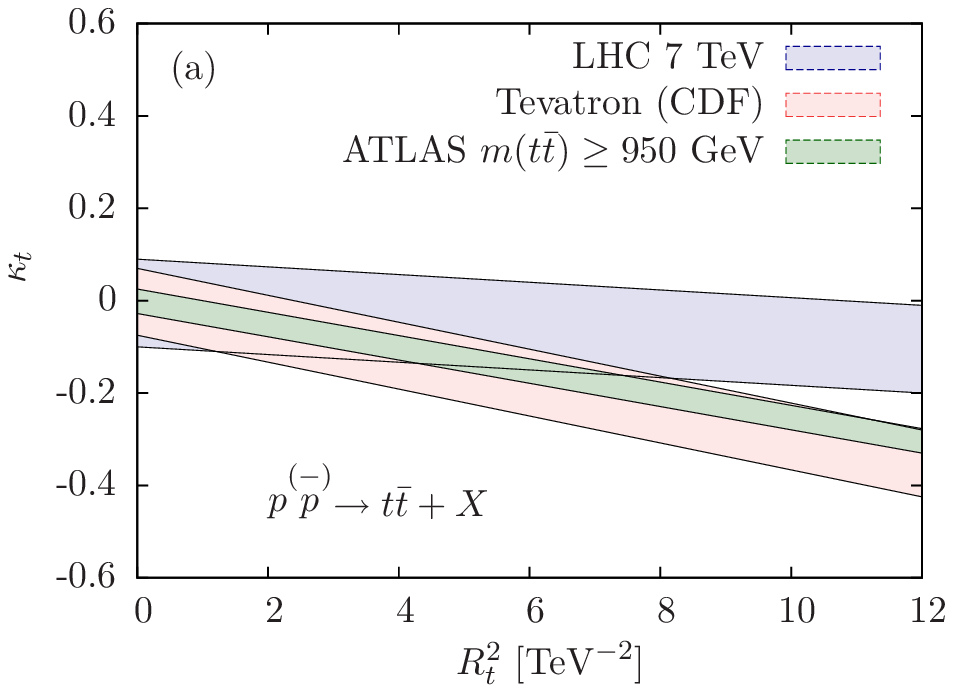}
  \hspace{1cm}
  \includegraphics[height=0.32\textwidth]{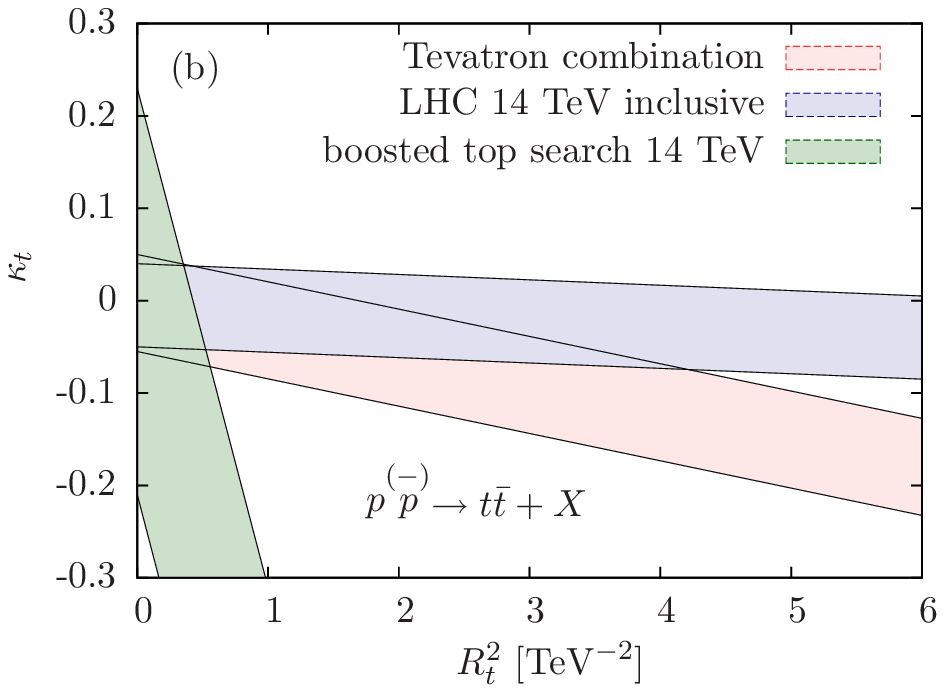}
  \caption{\label{fig:Rklines} {\it (a) Bands allowed in
      $R_t,\kappa_t$ space by $t\bar{t}$ production at Tevatron and
      LHC for 7 TeV, available data; (b) The same for LHC at 14 TeV,
      theoretical expectation of inclusive cross sections and boosted
      top events. We also show the region that is allowed by the high
      top pair invariant mass bin as reported in a recent ATLAS
      investigation of the differential $t\bar{t}$ cross section
      \cite{Aad:2012hg}, which is largely equivalent in sensitivity to
      the CDF analysis. We use $\mu_R=\mu_F=m_t$ and
      $\mu_F=\mu_R=\overline{m}_T$ for the boosted search, where
      $\overline{m}_T$ denotes the average transverse mass of the top
      quarks.}  }
\end{figure}

This procedure gives rise to a band of viable values in the
$\{R_t,\kappa_t\}$-parameter plane from each of the two colliders,
Fig.~{\ref{fig:Rklines}}. The crossing of the bands allows us to
determine the upper limits of the two parameters separately, resulting
in the conservative upper bounds collected in Tab.~{\ref{tab:Rk}}. At
the LHC, the inclusive $t\bar t$ cross section is driven by the gluon
fusion channel, which has no dependence on $R_t$, see Eq.~\gl{eq:2}.
This makes it difficult to obtain stringent bounds on $R_t$, in
contrast to the Tevatron where the quark-antiquark
channel is dominant, see Eq.~\gl{eq:1}. \\

The bound on $|\kappa|$ from the combination 'Tevatron $\oplus$ LHC[7
TeV]' of the presently available data would shrink to $|\kappa| <
0.06$ if the top radius is set to zero. Comparing this value with
appropriate values in the literature based on analyses of
chromo-magnetic and chromo-electric dipole moments
\cite{hioki,weiler}, they agree within errors of 30\%.\\

For the $t\bar t$ cross section at the Tevatron there are statistical
improvements upon combining the data sets of D\O\ and CD
\cite{tevcomb}. Similar improvements can be expected at the LHC for
the 14 TeV run, when more data will become available. We show a
projection of this situation in Fig.~{\ref{fig:Rklines}} (b), where we
scale the CDF error of Eq.~\gl{eq:deltasig} by a factor $1/\sqrt{2}$,
and the LHC systematic uncertainty is saturated at 5\%
\cite{searches}. This shows that we can indeed expect a significant
improvement on the limits of $\{R_t,\kappa_t\}$ at the level of
inclusive searches. \\

Even though we cannot carry out a rigorous analysis of quadratic
effects in the multipole expansion, nevertheless for LHC[7 TeV], as a
typical example, we may quote a rough estimate at what level the
linear term may penetrate the quadratic term. With $\langle s \rangle
\sim$ 1/4 TeV$^2$ the linear correction induced by the radius amounts
to $2 \langle s \rangle / \Lambda_\ast^2 \sim 0.7 < 1$, where however
it should be noted that this number is still reduced significantly by
a negative contribution of the magnetic term -- this destructive
interference being one of the crucial elements in our analysis. Taking
the estimate above at face value, the quadratic term inferred from the
interference term is less than about 10\%. Thus, the rough estimates
signal internal consistency
of our analysis. \\

However, despite its much larger collision energy, the sensitivity of
LHC to the anomalous top couplings is improved only moderately
compared to the Tevatron, as a result of the prevalence of the
$R_t$-insensitive gluon-fusion component in the total hadronic cross
section.  A way to eliminate this obstacle is to consider boosted top
final states \cite{boostedtop}. By restricting ourselves to large
momentum transfers we probe the incoming protons at large momentum
fractions, thus naturally shifting towards the $q \bar{q}$
contribution, which is more sensitive to $R_t$. This improvement more
than compensates for the significant reduction of the hadronic $t\bar
t$ production from imposing this cut.  We include this search channel
in Fig.~{\ref{fig:Rklines}}(b), where we choose $p_{T,t}\geq
1~{\rm{TeV}}$, for which we expect $\sigma_{\rm{SM}}\simeq
50~{\rm{fb}}$ and a 30\% measurement uncertainty. This error estimate
should be understood figuratively as a dedicated calculation in this
phase space region analogous to Refs.~\cite{moch,ttTH2,ttTH3,ttTH} is
currently not available. Recent analyses of the differential $t\bar t$
cross section \cite{Aad:2012hg} however suggest that this is roughly
the uncertainty that can be expected.  It is clear that by increasing
the transverse momentum selection, we probe larger partonic center of
mass energy, which in turn yields a larger sensitivity to the
anomalous parameters.  This stems from probing predominantly
quark-induced subprocesses at large partonic momentum fractions. We
can expect that the flat background distribution qualitatively behaves
$\sim p^{-2}_{T,\text{cut}}$ so that a background fluctuation is
parametrically described by $p^{-1}_{T,\text{cut}}$. On the other hand
the signal cross section in the dominant quark channels for the
boosted selection behave $\sim \Lambda^2_\ast$. Hence the sensitive
region for the boosted selection is characterized by $ \Lambda_\ast^2
\lesssim p_{T,\text{cut}}$ until the rate at a given luminosity is too
small to efficiently reconstruct the $t\bar t$ system. Thus, the
sensitivity will increase with the cut on the transverse momentum
until the error in the cross section becomes overwhelming.  By the
same reason, the precise value of the involved uncertainties is not
too important for the qualitative success of constraining the
anomalous top interactions using a boosted selection. It should also
be noted that these SM errors are expected to be also the main errors
in the part of the cross sections describing anomalous contributions.
Central sources for errors like the scales in the QCD coupling and the
parton densities are not significantly different from the SM,
i.e. they cancel out from observables like the radius, operatively
defined in a ratio of cross sections.  A combination of either
inclusive LHC cross sections together with finalized Tevatron results,
or inclusive cross sections and boosted searches solely at the LHC
provide good prospects
to sharpen the bounds on anomalous top interactions. \\

\begin{table}[tb]
  \begin{tabular} {| l | c | c |}
    \hline
    & $R_t$                                             & $\;\;|\kappa_t|\;\;$ \\
    \hline\hline
    $\;\;$Tevatron $\oplus$ LHC[7 TeV]    $\;\;$            &
    $\;\;$2.9 TeV$^{-1}$ $\sim$ $0.57 \times 10^{-16}$ cm $\;\;$& $\;\;0.17\;\;$ \\
    $\;\;$Tevatron $\oplus$ LHC[14 TeV]   $\;\;$            & 
    $\;\;$2.1 TeV$^{-1}$ $\sim$ $0.41 \times 10^{-16}$ cm $\;\;$& $\;\;0.07 \;\;$\\
    $\;\;$LHC[14 TeV]: inclusive $\oplus$ boosted top$\;\;$ & 
    $\;\;$0.7 TeV$^{-1}$ $\sim$ $0.14 \times 10^{-16}$ cm $\;\;$& $\;\;0.05 \;\;$\\
    \hline
  \end{tabular}
  \caption{\label{tab:Rk} {\it Upper bounds on $t$ radius and magnetic moment
      after combining Tevatron and LHC data / future expectations
      for $t\bar{t}$ production -- inclusive and boosted top measurements 
      at LHC.}} 
\end{table}

The anomalous parameters $R_t,\kappa_t$ can be translated to the scale
parameters $\Lambda_\ast$ and $\Lambda_\ast/\sqrt{|\rho|}$ [as denoted
in Eqs.~{\gl{eq:Rlam}} and {\gl{eq:klam}}].  Using the estimated
bounds on the radius $R_t$ from the Tevatron and the LHC experiments,
one obtains
\begin{equation}
  {\rm{Tevatron}} \;\oplus\; {\rm{LHC[7\; TeV]}} \;:\quad \Lambda_\ast
  \gtrsim 0.84 \;{\rm{TeV}}     \,,
\end{equation}
while the bound from the anomalous magnetic moment, for the
characteristic choice $\rho = 1$, is weaker by a factor
of~2. Identifying $g_s^2/\Lambda_*^2 \to c_{Vv}/(2\Lambda^2)$, this
number is in agreement with Ref.~\cite{Degrande:2010kt} when taking
into account the different conventions, and the bounds can be improved
by fitting the di-top invariant mass distribution
\cite{Degrande:2010kt}. \\

ATLAS has already published results on centrally produced, high
invariant-mass top pairs, $m_{t\bar t}\geq 950$~GeV
\cite{Aad:2012hg}. We find that limits obtained from this
result are compatible with the combined analysis 'Tevatron $\oplus$ LHC[7 TeV]'. \\

Improvements of the bound are expected for LHC[14 TeV], particularly
if boosted top analyses are exploited:
\begin{align}
&&&&
  &{\rm Tevatron \;\oplus\; LHC[14\; TeV]}\; : && \Lambda_\ast 
  \gtrsim 1.17 \;{\rm{TeV}}  \,, &&&& \\
&&&&
  &{\rm LHC[14\; TeV; \; inclusive \; \oplus \; boosted\; top]}
                           \;:  &&\Lambda_\ast \gtrsim 3.5 \;{\rm TeV}   
  \,, &&&&
\end{align}
dominated again within a factor of at least 2 by the bound on the
radius. 
Boosted strategies are not applicable at the Tevatron due the limited
data set and the small available center-of-mass energy compared to
LHC[14 TeV]. The LHC[7 TeV] data sample is also too small, but first
results can be expected from the LHC[8 TeV] run.\\

Including the measurements of angular distributions and spin
correlations in the experimental analyses will lift these limits to
still higher values. \\[2mm]

\noindent 
As mentioned above, the scale parameter $\Lambda_*$ can also be
transcribed to octet contact interactions, lifting the scale parameter
by half an order of magnitude, {\it cf.}
Eq.({\ref{Eq:ct}}). Presently a bound of $\Lambda_{\rm ct} \gtrsim
2.8$~TeV has been reached. The bound will improve significantly at
14-TeV LHC,
\begin{equation}
  \Lambda_{\rm ct} \gtrsim 11.7 \;\, {\rm TeV}    \,,
\end{equation}
in the near future, corresponding for singlet currents even to an
estimated 17.0 TeV for the singlet energy density identified,
hypothetically, with the color averaged octet density.\\[2mm]

\begin{figure}[!t]
  \includegraphics[height=0.32\textwidth]{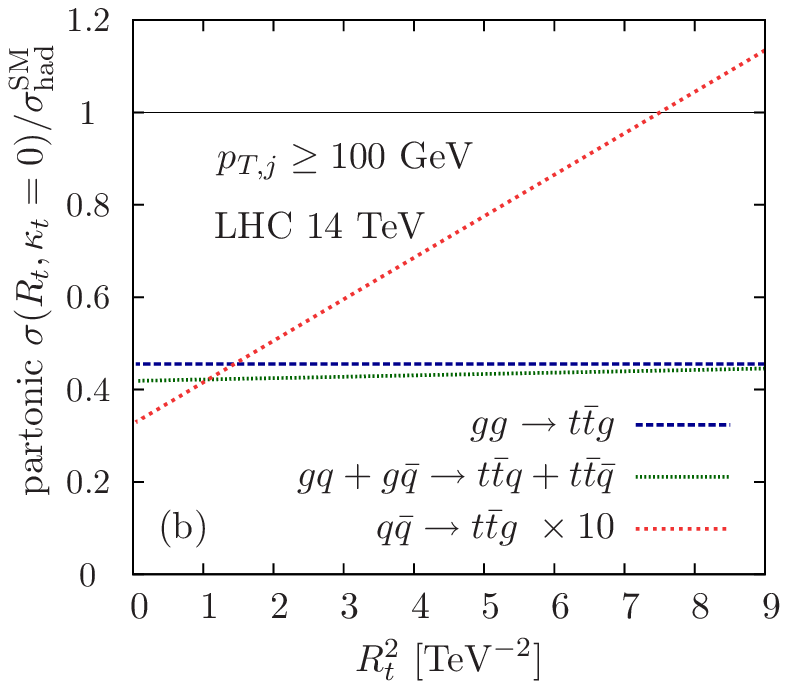} 
  \hspace{10mm}
  \includegraphics[height=0.32\textwidth]{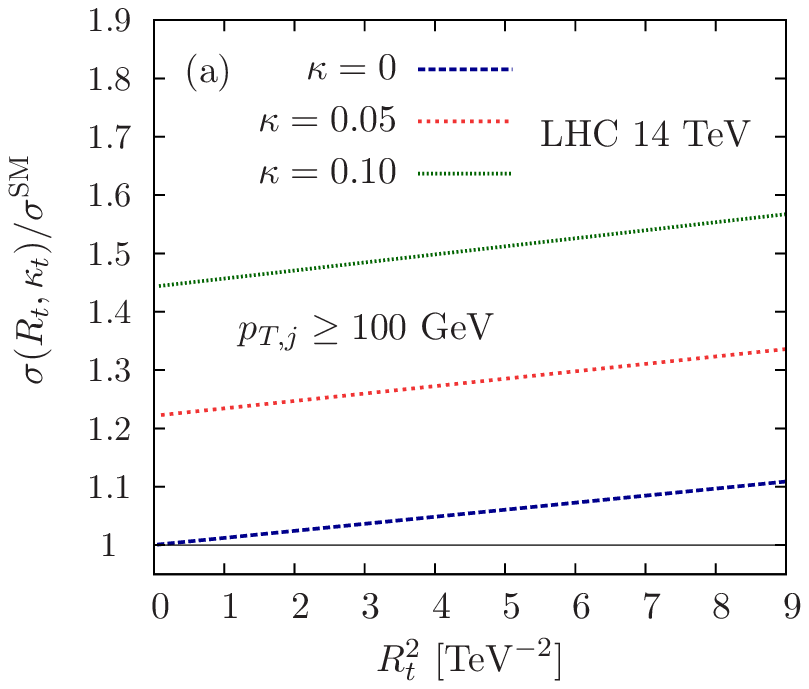}
  \caption{\label{fig:jet} {\it Dependence of the jet cross sections
      $\sigma[t\bar{t}j]$ on $R_t,\kappa_t$. Left panel: gluon fusion,
      (anti-)quark-gluon scattering, and quark-antiquark annihilation
      subprocesses compared with the original Born cross sections at
      the LHC $\sqrt{s}=14$ TeV; Right panel: LHC cross sections for
      various values of $\kappa_t$. We use again
      $\mu_R=\mu_F=\overline{m}_T$. } }
\end{figure}

\setcounter{section}{4}
\setcounter{equation}{0}
\noindent {\bf 4. Jet Emission.} Earlier we argued that the classical
Rutherford process $qt \to qt$ can be exploited for measuring the
radius of the $t$-quark while the Thomson analogue $gt \to gt$ does
not depend on the radius to leading order. These rules are also
effective in the crossed channels $q \bar{q} \to t \bar{t}$ and $gg
\to t \bar{t}$ applied in practice to measure the $t$ radius and the
magnetic moment.  Adding a gluon jet to the final state in $\sigma [gg
\to t \bar{t} g]$, the gluon-fusion process still has no dependence on
the radius. This is obvious for the logarithmically enhanced splitting
process $g \to gg$ followed by $gg \to t \bar{t}$, but it remains true
also for the non-logarithmic part. This is a consequence of
cancellations among the modified three-point ($gt\bar{t}$) vertex and
novel four- and five-point ($ggt\bar{t}$, $gggt\bar{t}$) vertex
contributions to $gg\to t\bar t g$ [resulting from
Eq.~\gl{eq:interactions} and \gl{eq:interactions2}], which do not only
serve to enforce the QCD Ward-identities but also eliminate the
$R_t$-dependent terms.  In Fig.~\ref{fig:jet}, it is demonstrated
numerically that indeed $\sigma [gg \to t \bar{t} g]$ is independent
of $R_t$. A cut, $p_{T,j}\geq 100$~GeV, has been imposed on the
transverse momentum of the jet. By contrast, the subprocess $gq \to t
\bar{t} q$ depends, weakly though, on the top radius already to
logarithmic accuracy through the gluon splitting channel $g \to q
\bar{q}$ followed by $q \bar{q} \to t \bar{t}$, supplemented by
additional non-logarithmic contributions.

However, this subchannel is dominated by gluon radiation $q \to qg$
and $gg \to t \bar{t}$, which depends on the top radius only beyond
the leading logarithmic order. The strongest $R_t$ dependence is
predicted for the annihilation channel $q \bar{q} \to t \bar{t} g$ via
two steps, $q \to qg$ and $q\bar{q} \to t \bar{t}$, with logarithmic
enhancement. The equivalent hadron cross section is shown in the right
panel of Fig.~\ref{fig:jet}.

Sensitivity to the radius in $t\bar t+{\rm{jet}}$ production is
largely driven by the $q\bar{q} \to t\bar{t}g$ subprocess, but some
additional sensitivity arises from $gq$ scattering (and the
charge-conjugated $g\bar q$ channel). As discussed above, the latter
originates from two contributions: \textit{(i)} splitting of an
initial-state gluon into a quark-antiquark pair, $g\to q\bar q^*$,
followed by the $R_t$-dependent subprocess $q\bar{q}^* \to t\bar{t}$;
\textit{(ii)} radiation of an off-shell gluon from the incident quark,
$q \to qg^*$, followed by $gg^* \to t\bar{t}$. To leading logarithmic
order, {\it{i.e.}} for nearly on-shell gluons, the second process is
independent of $R_t$ as argued earlier. However, if sufficiently
high-$p_T$ jets are observed, the events are pushed out of the DGLAP
regime.  Since the intermediate gluon in this case is
\textit{off-shell}, the contribution of the operator
\eqref{eq:interactions} is not forbidden by Ward identities. This
situation corresponds to the Rutherford-type scattering discussed on
page~\pageref{ruth}. One finds for the $R_t$-dependent contribution to
the cross section \textit{(ii)} in linear approximation, the
transverse part singled out for the sake of transparent illustration:
\begin{equation}
  \frac{d\Delta\sigma [gq \to t \bar{t} q]}{d\cos\theta^* dq^2} \propto
  \frac{R_t^2}{6}\bigl [ (1-\xi)(23+50\xi-9\xi^2) +
  4(1-\xi)(37+27\xi)m_t^2/s - 64(7+9\xi)m_t^4/s^2 ]/(1-\xi)^2  +\dots
\label{eq:off}
\end{equation}
[while absent in $gg \to t \bar{t} g$ due to the anomalous $(g)ggt\bar
t$ vertices according to Eq,.~(\ref{eq:interactions})]. Here $q^2 < 0$
is the 4-momentum transfer in the quark-line, {\it{i.e.}} the
virtuality of the off-shell gluon, and $\xi = \beta^2\cos^2\theta^*$,
where $\theta^*$ is the angle of the top-quark in the $t\bar{t}$ rest
frame. As follows from scaling, the size of the anomalous part of the
cross section is determined by the radius and it is independent of
$q^2$ for $|q^2| \ll R_t^{-2}$, fulfilled in all realistic
configurations. For the sake of clarity, the chromo-magnetic moment
$\kappa_t$ has not been included in \eqref{eq:off}, but the results in
Fig.~\ref{fig:jet} are based on both
the operators and all relevant diagrams with exact kinematics. \\

Therefore, with sufficient experimental precision, the top radius
$R_t$ can also be probed in $t\bar t+{\rm{jet}}$ in a unique way that
will help to discriminate its effect from the anomalous top magnetic
moment.
The $t\bar t+{\rm{jet}}$ process has a smaller cross section
$\sigma_{\rm{NLO}}^{\rm{SM}}(p_{T,j}\geq 50~{\rm{GeV}})=375~{\rm{pb}}$
\cite{Melnikov:2011qx} as compared to inclusive $t\bar t$
production. A measurement of $t\bar t+{\rm{jet}}$ is also more
involved from an experimental systematics point of view, and currently
there is no dedicated analysis available at the LHC that targets the
high $p_T$ regime [see Ref.~\cite{atlasn} for a first measurement of
inclusive $t\bar t+j$] Nonetheless,
sensitivity can also be gained in this channel using similar
strategies as discussed in Sec.~3.
\\[2mm]

\setcounter{section}{5}
\setcounter{equation}{0}
\noindent {\bf 5. Summary.} The intrinsic structure of the
top-quark can sensitively be probed at the Tevatron and LHC by setting
bounds on the color radius and the color magnetic dipole moment of the
particle. Values of
\begin{equation}
R_t \lesssim 1.4 \times 10^{-17}\;{\rm cm} \;\; {\rm and} \;\; |\kappa_t| \lesssim 0.05
\end{equation}
can be expected from LHC running in the near future. Present bounds,
combined with Tevatron results will improve by factors of 4 and 3,
respectively. These values can be mapped into effective scale
parameters
\begin{equation}
  \Lambda_\ast \gtrsim 3.5 \;{\rm TeV} \quad {\rm and}\quad \Lambda_{\rm ct}
  \gtrsim 11.7 \;{\rm TeV}  \,
\end{equation}
[and potentially even 17 TeV for singlet currents], 
strongly constraining the pointlike character of the top quark. \\[10mm]

\noindent {\uline{\it Acknowledgements}}: CE acknowledges funding by
the Durham International Junior Research Fellowship scheme. CE also
thanks the CERN Theory Group for hospitality during the time when this
work was completed. The work of AF is partially supported by the
National Science Foundation under grants PHY-0854782 and PHY-1212635.

\vskip 15mm
\setcounter{section}{1}
\setcounter{equation}{0}
\renewcommand{\theequation}{\Alph{section}.\arabic{equation}}
\renewcommand{\thesection}{\Alph{section}}
\noindent {\uline{\bf Appendix}}: A few examples should illustrate
the dependence of relevant helicity amplitudes on the anomalous top
parameters. Other helicity amplitudes are related by $\mathcal{P}$ and
$\mathcal{C}$ symmetries, and the exchange $\cos\theta \to
-\cos\theta$. Dimensional parameters are scaled in $E= \sqrt{s}/2$.
\\

\noindent
{\uline{\it quark-antiquark annihilation}}:
\begin{align}
\mathcal{M}(q^i_R{\bar q}^j_L \to t^k_L {\bar t}^l_L) &=
           \dfrac{K_q}{6m_t\sqrt{s}}[2m_t^2(6+sR_t^2)+3s\kappa_t]\sin\theta \\
  &= -\mathcal{M}(q^i_L{\bar q}^j_R \to t^k_L {\bar t}^l_L) =
   -\mathcal{M}(q^i_R{\bar q}^j_L \to t^k_R {\bar t}^l_R) =
   \mathcal{M}(q^i_L{\bar q}^j_R \to t^k_R {\bar t}^l_R), \nonumber \\[1ex]
\mathcal{M}(q^i_L{\bar q}^j_R \to t^k_L {\bar t}^l_R) &=
           \dfrac{K_q}{6}(1+\cos\theta)[6+sR_t^2+6\kappa_t] \\
  &= -\mathcal{M}(q^i_R{\bar q}^j_L \to t^k_R {\bar t}^l_L), \nonumber \\[1ex]
\mathcal{M}(q^i_R{\bar q}^j_L \to t^k_L {\bar t}^l_R) &= 
  -\mathcal{M}(q^i_L{\bar q}^j_R \to t^k_R {\bar t}^l_L)
 = \mathcal{M}(q^i_L{\bar q}^j_R \to t^k_L {\bar t}^l_R)
 [\cos\theta \leftrightarrow -\cos\theta].
\end{align}

\noindent
{\uline{\it gluon fusion}}:
\begin{align}      
\mathcal{M}(g^a_Lg^b_L \to t^k_L {\bar t}^l_L) &= 
 \frac{1}{2m_t\sqrt{s}} \begin{aligned}[t]\biggl [ &K_t
  \Bigl [\frac{1+\cos\theta}{1-\beta\cos\theta}\bigl \{
  4m_t^2(1+\beta-\beta\cos\theta) + s(1-\beta)(2+\beta-\beta\cos\theta)\kappa_t
   \bigr\} \Bigr ]
  \\ +& K_u \Bigl [\cos\theta \leftrightarrow -\cos\theta\Bigr ]
  \\ -& K_s \cos\theta [4m_t^2 + s(1-\beta)\kappa_t]
 \biggr ], \end{aligned} \\
 &= -\mathcal{M}(g^a_Rg^b_R \to t^k_R {\bar t}^l_R), 
 \nonumber \displaybreak[0] \\[1ex]
\mathcal{M}(g^a_Lg^b_L \to t^k_R {\bar t}^l_R) &= 
 -\mathcal{M}(g^a_Rg^b_R \to t^k_L {\bar t}^l_L) = 
\mathcal{M}(g^a_Lg^b_L \to
 t^k_L {\bar t}^l_L)[\cos\theta \leftrightarrow -\cos\theta, \,
  \beta \leftrightarrow -\beta], \displaybreak[0] \\[1ex]
\mathcal{M}(g^a_Lg^b_R \to t^k_L {\bar t}^l_L) &= 
 \frac{\beta}{2m_t\sqrt{s}} \Bigl [\frac{K_t}{1-\beta\cos\theta} +
  \frac{K_u}{1+\beta\cos\theta} \Bigr ] \sin^2\theta\,
  [4m_t^2+s\kappa_t], \\
  &= \mathcal{M}(g^a_Rg^b_L \to t^k_L {\bar t}^l_L)
   = -\mathcal{M}(g^a_Lg^b_R \to t^k_R {\bar t}^l_R)
   = -\mathcal{M}(g^a_Rg^b_L \to t^k_R {\bar t}^l_R), 
 \nonumber \displaybreak[0] \\[1ex]
\mathcal{M}(g^a_Lg^b_L \to t^k_L {\bar t}^l_R) &= \sin\theta
\biggl [ K_t\Bigl (1+\kappa_t\frac{2-\beta\cos\theta}{1-\beta\cos\theta}\Bigr ) 
  -K_u\Bigl (\cos\theta \leftrightarrow -\cos\theta \Bigr ) 
  -K_s [1+\kappa_t]
 \biggr ], \\
  &= \mathcal{M}(g^a_Rg^b_R \to t^k_L {\bar t}^l_R)
   = \mathcal{M}(g^a_Lg^b_L \to t^k_R {\bar t}^l_L)
   = \mathcal{M}(g^a_Rg^b_R \to t^k_R {\bar t}^l_L), 
 \displaybreak[0] \\
\mathcal{M}(g^a_Lg^b_R \to t^k_L {\bar t}^l_R) &= 
 -\beta \Bigl [ \frac{K_t}{1-\beta\cos\theta} +
   \frac{K_u}{1+\beta\cos\theta} \Bigr ] \sin\theta(1+\cos\theta)\,
  [1+\kappa_t], \\
  &= \mathcal{M}(g^a_Rg^b_L \to t^k_R {\bar t}^l_L), 
 \nonumber \displaybreak[0] \\[1ex]
\mathcal{M}(g^a_Rg^b_L \to t^k_L {\bar t}^l_R) &= 
 \mathcal{M}(g^a_Lg^b_R \to t^k_R {\bar t}^l_L) = 
 \mathcal{M}(g^a_Lg^b_R \to t^k_L {\bar t}^l_R)
  [\cos\theta \leftrightarrow -\cos\theta, \, \beta \leftrightarrow -\beta],
\end{align}

The color factors are defined as $K_q = g_s^2(T^a_{ij})^*T^a_{kl}$,
$K_s=ig_s^2f^{abc}T^c_{kl}$, $K_{t} =g_s^2T^a_{ki}T^b_{il}$, and
$K_{u} =g_s^2 T^a_{il} T^b_{ki}$, the evaluation of their products
follows the standard SU(3) rules, resulting in $K_q^\ast K_q = g_s^4N_cC_F/2$,
$K_s^\ast K_s = g_s^4N_c^2C_F$, $K_t^\ast K_t = K_u^\ast K_u = g_s^4N_cC_F^2$, 
and $K_s^\ast K_t = -K_s^\ast K_u = g_s^4N_c^2C_F/2$,
$K_t^\ast K_u = -g_s^4C_F/2$. \\



\end{document}